\documentclass[sigconf]{acmart}
\usepackage[T1]{fontenc}

\usepackage{multirow}
\usepackage{tcolorbox}
\usepackage{graphicx}
\usepackage[framemethod=TikZ,skipabove=6pt,skipbelow=6pt]{mdframed}
\usepackage{fontawesome5}
\usepackage{cleveref}
\usepackage{svg}
\usepackage{hyperref}
\usepackage{listings}
\usepackage{xcolor}
\usepackage{caption}
\usepackage{subcaption}

\AtBeginDocument{%
  }

\newcommand{\summarybox}[1]{   
\begin{mdframed}[hidealllines=true, backgroundcolor=gray!5]
{\textit{#1}}
\end{mdframed}
}

\setcopyright{acmlicensed}
\copyrightyear{2018}
\acmYear{2018}
\acmDOI{XXXXXXX.XXXXXXX}

\acmConference[E-QSE\@]{2nd edition of the workshop on Empirical Studies for Quantum Software Engineering}{June 2025}{Istanbul, Turkey}

\begin{document}

\title{A Preliminary Investigation on the Usage of Quantum Approximate Optimization Algorithms for Test Case Selection}

\author{Antonio Trovato}
\affiliation{
  \institution{University of Salerno}
  \city{Fisciano (SA)}
  \country{Italy}}
\email{atrovato@unisa.it}

\author{Martin Beseda}
\affiliation{
  \institution{University of L'Aquila}
  \city{L'Aquila}
  \country{Italy}}
\email{martin.beseda@univaq.it}

\author{Dario Di Nucci}
\affiliation{
  \institution{University of Salerno}
  \city{Fisciano (SA)}
  \country{Italy}}
\email{ddinucci@unisa.it}

\renewcommand{\shortauthors}{Trovato et al.}

\begin{abstract}
Regression testing is key in verifying that software works correctly after changes. However, running the entire regression test suite can be impractical and expensive, especially for large-scale systems.
Test suite optimization methods are highly effective but often become infeasible due to their high computational demands.
In previous work, Trovato et al. proposed \textsc{SelectQA}, an approach based on quantum annealing that outperforms the traditional state-of-the-art methods, i.e., \textsc{Additional Greedy} and \textsc{DIV-GA}, in efficiency.

This work envisions the usage of \textit{Quantum Approximate Optimization Algorithms} (QAOAs) for test case selection by proposing \textsc{QAOA-TCS}. QAOAs merge the potential of gate-based quantum machines with the optimization capabilities of the adiabatic evolution.
To prove the effectiveness of QAOAs for test case selection, we preliminarily investigate QAOA-TCS leveraging an ideal environment simulation before evaluating it on real quantum machines. Our results show that QAOAs perform better than the baseline algorithms in effectiveness while being comparable to SelectQA in terms of efficiency.
These results encourage us to continue our experimentation with noisy environment simulations and real quantum machines.
\end{abstract}

\ccsdesc[500]{Software and its engineering~Software testing and debugging}
\ccsdesc[500]{Hardware~Quantum computation}
\ccsdesc[500]{Software and its engineering~Search-based software engineering}

\keywords{Regression Testing, Quantum Computing, Search-based Software Engineering}


\maketitle

\section{Introduction and Background}

Software quality assurance is essential in modern software development, especially as systems undergo frequent updates and changes. Regression testing~\cite{yoo2012regression} is used to verify that previously tested software components continue functioning correctly and ensure these changes do not introduce defects. However, re-running all test cases after each update can be impractical and costly, particularly for large-scale systems. Therefore,  strategies such as test case selection (TCS) and test case prioritization (TCP) have been developed to reduce the testing burden while maintaining comprehensive coverage. However, these techniques often rely on complex optimization algorithms, which can be resource-intensive and difficult to scale.
Approaches based on quantum computing~\cite{nielsen2010quantum,horowitz2019quantum}, e.g., BootQA~\cite{bootqa}, IGDec-QAOA~\cite{igdec_qaoa}, and SelectQA~\cite{selectqa}, offer a potential solution to these limitations by leveraging its parallelism and exponential computational power to optimize test case selection more efficiently.

This work proposes QAOA-TCS, a new approach based on Quantum Approximate Optimization Algorithms (QAOA) for Test Case Selection, which combines the potential of gate-based approaches and leverages clustering techniques for efficient and data-driven decomposition. We developed a formulation of the problem coherent with Panichella et al.~\cite{panichella2014improving} and compared its effectiveness with classical methods, i.e., DIV-GA~\cite{panichella2014improving}, and another quantum-based approach, i.e., SelectQA~\cite{selectqa}. The latter was chosen for comparison since it was evaluated against the aforementioned classical approaches using the same software systems employed in this work.
We did not compare against other quantum-based methods, e.g., BootQA~\cite{bootqa} and IGDec-QAOA~\cite{igdec_qaoa}, since they require a different formalization of the test case selection problem and a much more in-depth analysis of the results. We leave the comparisons with these two approaches as future work.

This work is a preliminary analysis of QAOA's capabilities in solving the Multi-Objective Test Case Selection problem, running a QAOA-TCS leveraging an ideal quantum backend.
We use the \texttt{StatevectorSimulator}, a literal state vector or density matrix in a linear algebra sense, to obtain an exact simulation with no involved noise that applies the necessary rounding when performing floating-point operations. Our results show that QAOA-TCS outperforms traditional baselines and SelectQA in effectiveness. Nevertheless, QAOA-TCS and SelectQA are similar in terms of efficiency.

The paper is organized as follows. \Cref{sec:QAOA-TCS} introduces QAOA-TCS, our novel approach for test case selection based on Quantum Approximate Optimization Algorithms. \Cref{eval_traditional} details our research method, whose results are presented in \Cref{sec:results} and threats to validity in \Cref{sec:ttv}. Finally, \Cref{sec:conclusion} concludes the paper and provides future research directions.

\section{A New Approach for Test Case Selection based on Quantum Approximate Optimization Algorithms}\label{sec:QAOA-TCS}
The first step to solving the TCS problem using the QAOA approach~\cite{choi2019tutorial} consists of formulating the corresponding cost Hamiltonian. Since we aim to compare QAOA-TCS with classic and quantum algorithms, we developed a formulation of the problem coherent with Panichella et al.~\cite{panichella2014improving} and Trovato et al.~\cite{selectqa}. In both cases, the QUBO framework expresses the inclusion or exclusion of a test case in the final solution using the binary variable $x$.

In this three-objective formulation, we aim to find a sub-suite of the initial one that represents the optimal trade-off between execution cost, past faults coverage, and statement coverage.
To build the QUBO, we followed the methodologies by Glover et al.~\cite{qubo_tutorial}, by which, starting from a linear constrained binary optimization problem, we can derive the corresponding QUBO. Since the execution costs and past faults are bonded to single test cases, we formulate them in the linear function. The statement coverage is the quadratic constraint of the problem since distinct test cases can execute the same statement block, and we need to select just one to cover it. We move the quadratic constraint to the linear function using a penalty constant $P$, guiding the optimizer to acceptable solutions, i.e., solutions that do not incur penalties. To choose the value of $P$ that aligns with the application domain, we used the Upper Bound technique~\cite{penalty_weights}, which sets $P$ as slightly higher than the upper bound of the initial linear objective function.
The QUBO formulation is:

\begin{align}
H =&\ \alpha\sum_{i=1}^{|\Gamma|} x_i \cdot cost(\tau_i) \notag \\
  & -(1-\alpha)\sum_{i=1}^{|\Gamma|} (e_i \cdot x_i) \notag \\
  & + \sum_{k}\sum_{i,j \in T_k} (-P x_i + 2 P x_i x_j)
\end{align}

where $\alpha$ is a weight coefficient balancing the cost and past fault coverage objectives, $\Gamma$ is the starting test suite, $\tau_i$ is the $i-th$ test case, $k$ is the $k-th$ statement, $T_k$ the list of test cases executing it, and $e_i$ a boolean value expressing whether the $i-th$ test case has spotted a fault in the past.

Qubit availability is the most significant limitation of the currently available quantum computers.
Therefore, the test suites to optimize are decomposed into smaller sub-suites solvable by QAOA simulators using clustering techniques. Applying clustering allows for preserving the similarities between test cases in sub-suites, and since similar test cases tend to have redundant coverage, the optimization process will facilitate useless duplications. Also, the similarities between test cases of the same clusters imply a diversified representation of the initial suite, facilitating targeted balancing of the objective while building a diversified final sub-suite.
We perform K-Means clustering based on three normalized features of each test case: execution cost, past fault coverage, and statement coverage. These features are selected to capture key aspects influencing test case effectiveness and redundancy.
The number of clusters is determined manually for each program, considering the number of test cases and the hardware constraints, particularly the maximum number of test cases the QAOA simulator can handle. To ensure compatibility with these constraints, we further limit the size of each cluster to a maximum of 30 test cases. Excess test cases are reassigned to nearby clusters based on their distance in feature space.
The idea of applying a decomposition strategy to solve TCS using QAOAs has already been applied by Wang et al.~\cite{igdec_qaoa}. However, while IGDec-QAOA assigns an impact value to each test case and optimizes subsets iteratively, our method groups similar test cases before optimization, ensuring that internal redundancy is minimized. This distinction is critical: IGDec-QAOA iteratively selects impactful test cases based on their immediate effect on the objective function, whereas clustering preserves structural relationships within the test suite from the outset.

\section{Empirical Design}
\label{eval_traditional}
This section presents the design of our research method to assess the effectiveness and efficiency of the algorithms.

\paragraph{Goal and Research Questions.}
We evaluate the efficiency and effectiveness of the ideal QAOA-TCS by comparing it to the current classical state-of-the-art and quantum SelectQA approaches and assessing its applicability using real quantum machines.
To evaluate QAOA's robustness and effectiveness, we have tested it using the \texttt{StatevectorSimulator}, as described in \Cref{sec:QAOA-TCS}.
We compared the algorithms posing the following research questions:

\begin{itemize}
    \item \textbf{RQ1.} What is the \textit{effectiveness} of QAOA-TCS compared to traditional state-of-the-art and quantum SelectQA Test Case Selection techniques?
    \item \textbf{RQ2.} What is the \textit{efficiency} of QAOA-TCS compared to tradition state-of-the-art and quantum SelectQA Test Case Selection techniques?
\end{itemize}


\paragraph{Study Context.}
The \textit{study context} in this formulation consists of four GNU open-source programs from the \textit{Software-artifact Infrastructure Repository} (SIR)~\cite{sir_rep}: \texttt{flex}, \texttt{grep}, \texttt{gzip}, and \texttt{sed}.

\begin{table}[ht]
\centering
\footnotesize
\caption{Characteristics of the programs under study}\label{table:sir_programs}
\begin{tabular}{@{}lrrr@{}} 
\toprule
\textbf{Program} & \textbf{LOC} & \textbf{\# TCs} & \textbf{Description} \\
\midrule
flex    & 10,459   & 567  & Fast lexical analyzer  \\
grep    & 10,068   & 808  & Regular expression utility  \\
gzip    & 5,680    & 215  & Data compression program  \\
sed     & 14,427   & 360  & Non-interactive text editor  \\
\bottomrule
\end{tabular}
\end{table}

\Cref{table:sir_programs} summarizes the key characteristics of the four GNU open-source programs used in this study. These programs were selected because they represent a diverse range of code sizes and test suite characteristics, allowing for an examination of how optimization techniques perform under different conditions. Larger test suites pose a greater challenge in terms of optimization, as redundant test cases must be filtered out while preserving fault detection capability. Programs with smaller test suites require a more precise selection approach, as there is a smaller margin for eliminating test cases without impacting coverage.
The following describes the problem criteria and data collection.

\begin{itemize}
\item \textit{Statement Coverage.} We used \texttt{gcov}, which can track the statements executed by each test case in C programs.

\item \textit{Execution Cost.} The cost is calculated by counting the execution frequency of each basic block.

\item \textit{Past Faults Coverage.} The SIR provides \textit{fault matrices} to depict if a test case has detected errors in the past.
\end{itemize}

\paragraph{Experiment Design.}
The compared algorithms have been executed ten times to mitigate the randomness effects of quantum algorithms.
Since quantum annealing and QAOAs are single-objective algorithms, we built the Pareto fronts incrementally following the same strategy used by additional greedy by Yoo and Harman~\cite{yoo_harman}. We incrementally produced a set of non-dominated solutions from the selected test cases obtained by the optimization processes.

To answer RQ1, we compared the Pareto fronts obtained by the four algorithms to evaluate their \textit{effectiveness} coherently with Panichella et al.~\cite{panichella2014improving}. In particular, we built a reference Pareto front made by all the non-dominated solutions found by all the algorithms during all ten runs.
Let $P=\{P_1,...,P_l\}$ be the set of $l$ different Pareto fronts obtained after all the experiment runs with all the evaluated algorithms. The Pareto front $P_{ref}$ is defined as $P_{ref} \subseteq \bigcup_{i=1}^{l} P_i$ where $\forall p \in P_{ref} \nexists q \in P_{ref} : q > p$.
Given the reference front, we computed, as an effectiveness metric, the \textit{number of non-dominated solutions}, i.e., the number of non-dominated solutions found by an algorithm selected for the final reference front. 
To ensure the empirical reliability of the results for both effectiveness and efficiency metrics, we applied the \textit{Kruskal-Wallis H test}~\cite{h_test}, and \textit{Dunn's test}~\cite{dunn_test}, together with the \textit{Vargha-Delaney effect size} (\^A${12}$)~\cite{a12} to assess the magnitude of differences between compared sets of non-dominated solutions.
We selected these tests because they are non-parametric methods that do not assume normality in the data but only data independence, as in our case.

The \textit{Kruskal-Wallis H test} was first applied to determine if there was a statistically significant difference between all samples. If the null hypothesis (i.e., no statistically relevant difference between the samples) was rejected ($p$-value < 0.05), we then conducted \textit{Dunn's test} to identify which specific pairs of samples exhibited significant differences. To control the false discovery rate, we employed Benjamini-Hochberg correction~\cite{benjamini1995controlling}, thus obtaining adjusted p-values. For p-values $<$0.05, we computed the \textit{Vargha-Delaney effect size} (\^A$_{12}$) to estimate the magnitude of the difference between the identified couples of samples.

Regarding RQ2, we analyzed the total execution times required by the algorithms. The runtime of QAOA-TCS consists of the execution time of the simulator launched on a supercomputer featuring an Intel Xeon Gold 6240 processor of 2.60GHz and allocating eight cores. For SelectQA, executed on the D-Wave \texttt{hybrid\allowbreak\_binary\allowbreak\_quadratic\allowbreak\_model\allowbreak\_version2}~\cite{tech_rep}, we considered the D-Wave API "total run time" metric; the other algorithms' execution times were computed on the same hardware used for QAOA-TCS.

\paragraph{Data Replication.}
All data and implementations are available in our online appendix~\cite{appendix}.

\section{Empirical Results}
\label{sec:results}
This section describes the results of both research questions.

\begin{table}[ht]
    \centering
    \scriptsize 
    \caption{Means and Standard Deviations of the Pareto Size and Number of Dominated Solutions for all Algorithms considered in our Empirical Study}
    \begin{tabular}{llrrrr}
        \toprule
        \multirow{2}{*}{Program} & \multirow{2}{*}{Method} & \multicolumn{2}{c}{Pareto Size} & \multicolumn{2}{c}{Non-Dom Solutions} \\
        \cmidrule(r){3-4} \cmidrule(r){5-6}
        &  & Mean & St. Dev. & Mean & St. Dev. \\
        \midrule
        flex & SelectQA & 173.0 & - & 173.0 & - \\
        & DIV-GA & 140.0 & - & 138.1 & 1.91 \\
        & Add. Greedy & \textbf{567.0} & - & 89.0 & - \\
        & QAOA-TCS & 447.2 & 17.45 & \textbf{408.2} & 23.78 \\
        \midrule
        grep & SelectQA & 229.5 & 0.5 & 3.5 & 0.52 \\
        & DIV-GA & 70.0 & - & 68.1 & 2.28 \\
        & Add. Greedy & \textbf{802.0} & - & 97.0 & - \\
        & QAOA-TCS & 481.4 & 12.33 & \textbf{472} & 14.34 \\
        \midrule
        gzip & SelectQA & 86.3 & 0.8 & 2.0 & - \\
        & DIV-GA & 105.0 & - & 64.0 & 24.06 \\
        & Add. Greedy & \textbf{199.0} & - & 62.0 & - \\
        & QAOA-TCS & 128.5 & 6.59 & \textbf{67.1} & 28.14 \\
        \midrule
        sed & SelectQA & 131.0 & - & 66.0 & - \\
        & DIV-GA & 99.6 & 13.4 & 93.3 & 13.07 \\
        & Add. Greedy & \textbf{356.0} & - & 75.0 & - \\
        & QAOA-TCS & 216.4 & 9.95 & \textbf{116.5} & 8.68 \\
        \bottomrule
    \end{tabular}
    \label{table:statevector_results}
\end{table}


\begin{table}[ht]
    \centering
    \footnotesize
    \caption{Kruskal-Wallis H Test Results for the Number of Non-Dominated Solutions}
    \begin{tabular}{llrr}
        \toprule
        \textbf{Program} & \multicolumn{2}{c}{\textbf{Non-Dom. Solutions}} \\
        \cmidrule(r){2-3}
        & \textbf{p-value} & \textbf{X$^2$(3)} \\
        \midrule
        flex & \textbf{$<$0.01} & 37.794 \\
        grep & \textbf{$<$0.01} & 37.342 \\
        gzip & \textbf{$<$0.01}  & 23.524 \\
        sed  & \textbf{$<$0.01} & 33.096 \\
        \bottomrule
    \end{tabular}
    \label{table:kw_non_dom}
\end{table}

\begin{table}[ht]
    \centering
    \footnotesize
    \caption{Dunn's Test with Benjamini-Hochberg Correction and \^A$_{12}$ for the Number of Non-Dominated Solutions}
    \begin{tabular}{llrrr}
        \toprule
        \multirow{2}{*}{\textbf{Program}} & \multirow{2}{*}{\textbf{Hypothesis}} & \multicolumn{2}{c}{\textbf{Non-Dom Solutions}} & \multirow{2}{*}{\textbf{\^A$_{12}$}} \\
        \cmidrule(r){3-4}
         & & \textbf{p-value} & \textbf{adj p-value} & \\
        \midrule
        \multirow{3}{*}{flex} & QAOA-TCS$>$DIV-GA & \textbf{$<$0.01} & \textbf{$<$0.01} & 1.0 (L) \\
        & QAOA-TCS$>$SelectQA &  0.05 & 0.08 & - \\
        & QAOA-TCS$>$Add. Greedy & \textbf{$<$0.01} & \textbf{$<$0.01} & 1.0 (L) \\
        \midrule
        \multirow{3}{*}{grep} & QAOA-TCS$>$DIV-GA &  \textbf{$<$0.01} & \textbf{$<$0.01} & 1.0 (L) \\
        & QAOA-TCS$>$SelectQA & \textbf{$<$0.01} & \textbf{$<$0.01} & 1.0 (L) \\
        & QAOA-TCS$>$Add. Greedy & 0.05 & 0.06 & - \\
        \midrule
        \multirow{3}{*}{gzip} & QAOA-TCS$>$DIV-GA &  0.48 & 0.58 & - \\
        & QAOA-TCS$>$SelectQA & \textbf{$<$0.01} & \textbf{$<$0.01} & 1.0 (L) \\
        & QAOA-TCS$>$Add. Greedy & 0.43 & 0.65 & - \\
        \midrule
        \multirow{3}{*}{sed} & QAOA-TCS$>$DIV-GA &  \textbf{$<$0.01} & \textbf{$<$0.01} & 0.96 (L) \\
        & QAOA-TCS$>$SelectQA & \textbf{$<$0.01} & \textbf{$<$0.01} & 1.0 (L) \\
        & QAOA-TCS$>$Add. Greedy & \textbf{$<$0.01} & \textbf{$<$0.01} & 0.9 (L) \\
        \bottomrule
    \end{tabular}
    \label{table:dunn_bh_non_dom}
\end{table}

\paragraph{RQ1 - Effectiveness.} 
\Cref{table:statevector_results} shows that QAOA-TCS consistently achieves the highest mean number of non-dominated solutions across all four programs. In other words, QAOA-TCS generates Pareto frontiers containing a larger proportion of non-dominated solutions across all executions, suggesting superior effectiveness in conducting test case selection. The statistical results supporting these conclusions are presented in \Cref{table:kw_non_dom,table:dunn_bh_non_dom}. The Kruskal-Wallis test indicated that all differences among the tested samples were statistically significant, with p-values lower than 0.01. The post-hoc analysis using Dunn's test confirmed that significant differences, again with p-values below 0.01, were observed in 8 out of 12 pairwise comparisons. These results were further reinforced by the Benjamini-Hochberg correction, which maintained the significance of the findings after controlling for multiple comparisons. The Vargha-Delaney statistic (\^A$_{12}$) was computed to provide an interpretable measure of effect size, with all values exceeding 0.5, indicating a meaningful distinction between the compared groups. Notably, 8 out of 12 \^A$_{12}$ values were equal to 1, emphasizing the strength of the observed differences.
The variations in algorithm performance between programs can be attributed to their differing characteristics. QAOA-TCS performed better than the other algorithms on \texttt{flex}, \texttt{grep}, and sed. This result suggests that QAOA-TCS is particularly advantageous for programs with medium, large, and diverse test suites, as it effectively selects non-dominated test cases across multiple runs. However, even in smaller test suites, it remains competitive, although there are no strong differences with the other approaches.

\summarybox{\faKey \hspace{0.2mm} \textbf{RQ1 Summary.}
 The advantage of QAOA-TCS in effectiveness is not always statistically conclusive, suggesting that while it generally performs better, further investigation may be required in specific contexts.
}

\begin{table}[ht]
    \centering
    \scriptsize 
    \caption{Means and Standard Deviations of the Execution Times for all Algorithms considered in our Empirical Study}
    \begin{tabular}{llrr}
        \toprule
        Program & Method & Mean & St.Dev. \\
        \midrule
        flex & SelectQA & \textbf{2.99s} & 0.03s \\
        & DIV-GA & 3m4s & 14.68s \\
        & Add. Greedy & 8.33s & $<$0.01s \\
        & QAOA-TCS & 18.66s & 1m20s \\
        \midrule
        grep & SelectQA & \textbf{2.99s} & $<$0.01s \\
        & DIV-GA & 1m28s & 5.22s \\
        & Add. Greedy & 8.88s & $<$0.01s \\
        & QAOA-TCS & 1m25s & 55.24s \\
        \midrule
        gzip & SelectQA & 2.98s & $<$0.01s \\
        & DIV-GA & 19.40s & 2.21s \\
        & Add. Greedy & 0.23s & 0.22s \\
        & QAOA-TCS & \textbf{0.07s} & 0.06s \\
        \midrule
        sed & SelectQA & 2.99s & $<$0.01s \\
        & DIV-GA & 1m19s & 8.07s \\
        & Add. Greedy & 1.89s & $<$0.01s \\
        & QAOA-TCS & \textbf{0.07s} & 0.02s \\
        \bottomrule
    \end{tabular}
    \label{table:runtime_results}
\end{table}


\begin{table}[ht]
    \centering
    \footnotesize
    \caption{Kruskal-Wallis Test Results for the Execution Times}
    \begin{tabular}{llrr}
        \toprule
        \textbf{Program} & \multicolumn{2}{c}{\textbf{Non-Dom. Solutions}} \\
        \cmidrule(r){2-3}
        & \textbf{p-value} & \textbf{X$^2$(3)} \\
        \midrule
        flex & \textbf{$<$0.01} & 84.750 \\
        grep & \textbf{$<$0.01} & 87.089 \\
        gzip & \textbf{$<$0.01}  & 87.341 \\
        sed  & \textbf{$<$0.01} & 86.700 \\
        \bottomrule
    \end{tabular}
    \label{table:kw_non_dom_time}
\end{table}

\begin{table}[ht]
    \centering
    \footnotesize
    \caption{Dunn's Test with Benjamini-Hochberg Correction and \^A$_{12}$ for the Execution Times}
    \begin{tabular}{llrrr}
        \toprule
        \multirow{2}{*}{\textbf{Program}} & \multirow{2}{*}{\textbf{Hypothesis}} & \multicolumn{2}{c}{\textbf{Execution Time}} & \multirow{2}{*}{\textbf{\^A$_{12}$}} \\
        \cmidrule(r){3-4}
         & & \textbf{p-value} & \textbf{adj p-value} & \\
        \midrule
        \multirow{3}{*}{flex} & SelectQA$<$DIV-GA & \textbf{$<$0.05} & \textbf{$<$0.01} & 0.0 (L) \\
        & SelectQA$<$QAOA-TCS & \textbf{$<$0.01} & \textbf{$<$0.05} & 0.0 (L) \\
        & SelectQA$<$Add. Greedy & \textbf{$<$0.05} & 0.07 & 0.0 (L) \\
        \midrule
        \multirow{3}{*}{grep} & SelectQA$<$DIV-GA & \textbf{$<$0.01} & \textbf{$<$0.01} & 0.0 (L) \\
        & SelectQA$<$QAOA-TCS & \textbf{$<$0.01} & \textbf{$<$0.01} & 0.0 (L) \\
        & SelectQA$<$Add. Greedy & 0.39 & 0.44 & 0.0 (L) \\
        \midrule
        \multirow{3}{*}{gzip} & QAOA-TCS$<$DIV-GA & \textbf{$<$0.01} & \textbf{$<$0.01} & 0.0 (L) \\
        & QAOA-TCS$<$SelectQA & \textbf{$<$0.01} & \textbf{$<$0.01} & 0.0 (L) \\
        & QAOA-TCS$<$Add. Greedy & \textbf{$<$0.05} & \textbf{$<$0.05} & 0.0 (L) \\
        \midrule
        \multirow{3}{*}{sed} & QAOA-TCS$<$DIV-GA & \textbf{$<$0.01} & \textbf{$<$0.01} & 0.0 (L) \\
        & QAOA-TCS$<$SelectQA & \textbf{$<$0.01} & \textbf{$<$0.01} & 0.0 (L) \\
        & QAOA-TCS$<$Add. Greedy & \textbf{$<$0.01} & \textbf{$<$0.01} & 0.0 (L) \\
        \bottomrule
    \end{tabular}
    \label{table:dunn_bh_non_dom_time}
\end{table}

\paragraph{RQ2 - Efficiency.} \Cref{table:runtime_results} shows the mean execution time and standard deviation for each considered method across the four programs.
SelectQA achieves the lowest and most stable execution times across all programs, maintaining a mean of approximately 2.99s with a negligible standard deviation.
Additional Greedy also exhibits strong efficiency, with a minimal standard deviation, but has a computational time of $O(|T| \cdot max|T_i|)$, $|T|$ being the size of the starting test suite and $max|T_i|$ the cardinality of the largest suite able to reach the maximum coverage. Larger $max|T_i|$ means a higher number of iterations, which makes additional greedy impractical for larger systems.
DIV-GA incurs significantly higher computational costs due to its evolutionary nature and diversity-injecting mechanisms, which require multiple iterations over a population of solutions.
QAOA-TCS demonstrates a distinct efficiency pattern. For \texttt{gzip} and sed, it achieves remarkably low execution times, whereas for \texttt{flex} and \texttt{grep}, its execution time increases dramatically. The high standard deviations in these cases suggest that the quantum-inspired optimization process exhibits considerable variability in convergence time when handling larger test cases.

The results were further investigated using the statistical tests as illustrated in \Cref{table:kw_non_dom_time,table:dunn_bh_non_dom_time}. The Kruskal-Wallis test rejected the possibility of similarity between the execution times with all p-values < 0.01. The post-hoc analysis via Dunn's test rejected it in 11 out of 12 cases, while after the Benjamini-Hochberg correction, 10 out of 12 cases were rejected. In all cases, the \^A$_{12}$ was $\sim0$.

The results suggest that QAOA-TCS and DIV-GA scale up less efficiently with test suite size. Conversely, \texttt{gzip} and sed benefit from exceptionally low execution times with QAOA-TCS, reinforcing that it is highly efficient when the optimization space is constrained.

\summarybox{\faKey \hspace{0.2mm} \textbf{RQ2 Summary.}
QAOA-TCS demonstrates competitive execution times for smaller test suites, but its efficiency significantly declines for larger ones. Statistical analysis confirms these differences. These results highlight the need for further optimizations and analysis, especially using real quantum machines.
}
    
\section{Threats to Validity}
\label{sec:ttv}

This section discusses the threats to the validity of our study.

\paragraph{Construct Validity}
One of the main threats to construct validity is the correctness of the selection criteria for optimization, including coverage, historical failure data, execution cost, and failure rate. We mitigated this problem thanks to the use of well-established tools for data extraction. Additionally, while QAOA-TCS relies on \texttt{statevector} simulation, which provides an idealized execution model, real quantum hardware could introduce noise affecting performance, which we will explore in further studies. Another potential issue arises from implementation differences. DIV-GA was developed using MATLAB, whereas Additional Greedy, SelectQA, and QAOA-TCS were implemented using Python.
These differences can introduce important variations regarding language efficiency and optimization mechanisms. The implementations were designed to follow the configurations of the algorithms as presented in~\cite{panichella2014improving} and \cite{selectqa}.
Another threat is that SelectQA has been executed on different hardware that uses real QPUs.

\paragraph{Internal Validity}
A major internal threat is the inherently stochastic nature of quantum-inspired and evolutionary algorithms. To mitigate this threat, all experiments were repeated ten times per program. QAOA-TCS, DIV-GA, and SelectQA involve hyperparameter tuning, which could affect their performance. We configured SelectQA and DIV-GA as in their original works and tested QAOA-TCS with different configurations to find the best one.
A potential threat is the choice of the number of clusters in K-Means, which may influence the optimization outcome. We mitigated this by validating the parameter through repeated empirical trials.

\paragraph{External Validity}
A potential limitation is the generalizability of the results. The study analyzes four GNU programs from the SIR repository, which, while representative of different software functionalities and test suite sizes, may not fully capture the complexity of large-scale industrial systems. However, these datasets have been widely used in prior work on test optimization, making them a reasonable benchmark.

\paragraph{Conclusion Validity}
All conclusions are based solely on statistically significant results. The assumptions of similarity were based on knowledge of the problem properties. However, for higher reliability of the conclusions, it would be necessary to perform rigorous similarity comparisons between tested distributions, e.g., combining Cram\' {e}r-von-Mieses,~\cite{cramer1928distribution}, Anderson-Darling,~\cite{anderson1952statistical} and Kolmogorov-Smirnov,~\cite{kolmogorov1933sulla} to see, whether the distributions are significantly different, accompanied by Levene's test,~\cite{levene1960robust} to check for the significance of difference in the distributions' variances. Finally, the visual comparison via histograms of Q-Q plots would be helpful for larger statistical tests and more in-depth conclusions.

\section{Conclusions}
\label{sec:conclusion}
This paper proposes QAOA-TCS, a new algorithm to perform Test Case Selection that leverages gate-based quantum computers.
Our preliminary investigation shows that QAOA-TCS is more effective in selecting test cases than state-of-the-art algorithms in finding better solutions.
QAOA-TCS exhibited mixed performance in terms of efficiency. Its execution times were low for smaller test suites and higher for larger ones. Statistical tests confirmed these results. However, we expect significantly better results when executing QAOA-TCS on real quantum machines.

Overall, QAOA-TCS proves to be a promising approach for regression test case selection. It achieves superior effectiveness while maintaining competitive efficiency. Future work will further investigate the performance of QAOA in solving the Test Case Selection problem, first considering quantum noisy environments and then executing it on real quantum machines.

\begin{acks}
  Finanziato dall’Unione Europea - Next Generation EU, Missione 4 Componente 1 CUP D53D23017570001 e E53D23016370001 e dal Governo italiano (Ministero dell'Università e della Ricerca, PRIN 2022 PNRR) -- cod. P2022SELA7: ``RECHARGE: monitoRing, tEsting, and CHaracterization of performAnce Regressions'' -- Decreto Direttoriale n. 1205 del 28/7/2023.
\end{acks}

\bibliographystyle{ACM-Reference-Format}
\bibliography{main.bib}

\end{document}